\documentclass[aps,pra,showpacs,notitlepage]{revtex4-1}

\usepackage{graphicx}

\begin{document}

\title{Dynamics and statistics in the operator algebra of quantum mechanics}

\author{Holger F. Hofmann}
\email{hofmann@hiroshima-u.ac.jp}
\affiliation{
Graduate School of Advanced Sciences of Matter, Hiroshima University,
Kagamiyama 1-3-1, Higashi Hiroshima 739-8530, Japan}

\begin{abstract}
Physics explains the laws of motion that govern the time evolution of observable properties and the dynamical response of systems to various interactions. However, quantum theory separates the observable part of physics from the unobservable time evolution by introducing mathematical objects that are only loosely connected to the actual physics by statistical concepts and cannot be explained by any conventional sets of events. Here, I examine the relation between statistics and dynamics in quantum theory and point out that the Hilbert space formalism can be understood as a theory of ergodic randomization, where the deterministic laws of motion define probabilities according to a randomization of the dynamics that occurs in the processes of state preparation and measurement.
\end{abstract}

\maketitle

\section{Introduction}
\label{sec:intro}

Quantum theory is unique in the history of science. No other theory of natural phenomena has caused as much confusion about the relation between logical concepts and experimental observations. It may therefore be necessary to take a step back and examine the reasons for the confusion without hastily committing to one of the many ideological camps that have sprung up in the course of the scientific discussion. To do so, we should remind ourselves that the scientific method is to resolve controversies by a direct appeal to shared experience in the form of experimental observations. If quantum theory is really a scientific theory, all controversies can be decided by focusing the discussion on the experimental evidence. Specifically, we need to take care that all our statements, no matter how sophisticated or abstract, can be fully explained in terms of their relevance for possible experimental observations. 

The historical problem of quantum theory is that it was developed with a minimum of experimental input, using extrapolations that were motivated mostly by the beauty of the mathematical formalism \cite{Heisenberg}. However, technology has advanced to the point where we can finally control and measure individual quantum systems. Interestingly, the effects we can now observe are correctly described and predicted by the original formalism, and yet we have not been able to resolve the paradoxes associated with quantum theory. Indeed, the number of paradoxes has only increased \cite{Bell,KS,LGI,Aha,box,Har,Che}, and many of these paradoxes have been confirmed experimentally without providing any hint of an underlying physical reality \cite{Res04,Jor06,Tol07,Wil08,Lun09,Yok09,Gog11,Suz12,Den14}. At the heart of all of these perplexing paradoxes lies the fact that we do not understand the quantum processes used to measure and control the physical properties of quantum systems. It is here where a proper revision of quantum mechanics should start: how does the established formalism deal with the problem of measurement and control? 

An interesting contribution to this important question has been provided by Ozawa, who showed that the uncertainties of quantum measurements can be much lower than textbook formulations of the uncertainty principle suggest \cite{Oza03}. Most importantly, this result was derived entirely from the algebraic structure of the Hilbert space formalism, without any speculations about the underlying realities. Experimental studies are possible and have been realized \cite{Has12,Roz12,Bae13,Wes13,Rin14}, but these methods rely on indirect evaluations of the uncertainties, illustrating the fundamental problem that it is impossible to obtain the uncertainty free value of the target observable in conjunction with the uncertain outcome of an individual measurement. The dilemma of quantum measurements is that one cannot go back in time and obtain the value of a different observable for the same system. In Ozawa's theory, the problem is solved by using the operator formalism to define the value of a physical property mathematically, but critics of this approach tend to insist on definitions of uncertainties that are based only on the experimentally observable statistics of measurement outcomes - a notion that is extremely restrictive in the context of quantum mechanics \cite{Wat11,Bus13,Dre14,Bus14}. 

It seems to me that the present discussions are missing the actual point. Clearly, Ozawa's theory is valid within the stage set by the formalism. The confusion arises because the self-adjoint operators used to describe physical properties cannot be identified with the measurement outcomes through which we experience the physical property. To solve this problem, we need to review why quantum theory seems to introduce physical properties in two different and essentially incompatible ways - both as qualitative measurement outcomes with possible statistical errors and as quantitative shifts of pointer position averages associated with the external measurement setup (the ``meter system''). In the formalism, this dualism between quality and quantity is represented by operators, with the measurement operators of positive valued operator measures (POVMs) describing the qualitative outcome and the self-adjoint operators associated with observable properties of the system describing the quantity that is responsible for the pointer shift of the meter \cite{Nii16}. As I will show in the following, the problem can be addressed by considering the peculiar role of unitary dynamics in the formalism, which leads to a new understanding of the action in quantum statistics \cite{Hof16}. From the mathematical side, the close relation between unitary transformations and self-adjoint operators established by their corresponding eigenstates indicates that the eigenstate projectors represent time-averaged orbits of the dynamics, and not just the selection of a specific subset from a set of pre-determined realities. In the theoretical description of a quantum measurement, the dephasing processes associated with the observation of a precise outcome correspond to a dynamical randomization along the complete orbit. Importantly, it is not possible to separate this ergodic orbit into individual phase space points. Both the experimental evidence and the theoretical description therefore suggest that each measurement samples the complete dynamics generated by the target observable.  

In this paper, I will explain the relation between the elements of Hilbert space algebra and the experimental processes used in the laboratory. It is then possible to see that the algebra describes the fundamental laws of physics that govern physical interactions at the absolute limit of control set by the fundamental constant $\hbar$. In particular, I will address the origin of probabilities and the reason why quantum statistics is different from classical phase space statistics. The central result is that our understanding of experiments and experimental evidence cannot be based on preconceived notions of reality, but should instead emerge from the laws of causality that relate phenomena to physical objects. Quantum mechanics only appears strange and confusing because we fail to include the role of the dynamics in these causality relations. At the order of magnitude defined by the constant $\hbar$, Hilbert space is needed to express the dynamical structure of physical processes, which is more fundamental than the cruder notion of static realities commonly used in classical physics. 

\section{The physics of Hilbert space}
\label{sec:physics}

Many introductions to quantum mechanics start from the assumption that physical systems are described by a ``state''. The problem with such an introduction is twofold. Firstly, real systems are usually in motion, and secondly, the word ``state'' has no meaning until we explain how the ``state'' can describe a specific situation found in the real world. Interestingly, the closest practical analogy to the use of the term ``state'' in quantum theory is found in statistical physics, where thermal states are described by ergodic averages of their motion, with each orbit obtaining a statistical weight according to the energy of the orbit. In fact, we can see that the analogy works perfectly in quantum mechanics, where the thermal state is given by the density operator
\begin{equation}
\label{eq:thermal}
\hat{\rho}=\sum_n \frac{1}{Z}\exp(-\frac{E_n}{k_B T}) \mid \psi_n \rangle \langle \psi_n \mid.
\end{equation} 
The canonical partition function $Z$ is defined as in classical physics and the projectors on the energy eigenstates $\mid \psi_n \rangle$ take over the role of the orbits of energy $E_n$. 

Thermal states are time independent by definition. In quantum theory, this is particularly easy to see, since the energy eigenstates are also eigenstates of the unitary transformation $\hat{U}(t)$ that describes the time evolution of states. In fact, the similarity between the theoretical representation of time evolution and the representation of time independent ergodic states is a non-trivial feature of quantum theory that should not be underestimated. I hope that the arguments I am presenting here will draw more attention to this fact, and to the necessary consequences for our understanding of physics. Specifically, the time evolution is represented by an operator of the form
\begin{equation}
\label{eq:unitary}
\hat{U}(t) = \sum_n \exp(-i\frac{S_n}{\hbar}) \mid \psi_n \rangle \langle \psi_n \mid,
\end{equation}
where the action $S_n$ is given by the product of energy and time, $E_n t$. Two observations are important here. Firstly, no such operator PAPERS exists in classical physics, and this makes it extremely difficult to identify the actual relations between classical concepts and the Hilbert space algebra. Secondly, the action $S_n$ is the quantity that defines the amount of change induced by $\hat{U}$, and it is here that the fundamental constant $\hbar$ obtains a physical meaning. 

Experimentally, we can control systems by manipulating their interactions using the available forces, very often in the form of rather strong electromagnetic fields. Unfortunately, most systems are also experiencing a wide range of completely uncontrolled interactions, and this often limits the quality of control to a level where quantum effects cannot be seen. Note that the presence of these uncontrolled interactions means that the mathematical structure of classical physics is not confirmed by any experimental results, since the correspondence between experimental result and classical theory is merely an approximate fit valid at very limited resolution. Differential equations are only successful in describing real world physics because their solutions roughly approximate those patterns in our experience of nature that are robust to the extra noise of real life physics. To investigate the actual laws of physics, we need to remove these extra noise sources, and that is quite difficult. In many cases, it involves vacuum chambers and highly specialized methods of cooling. 

To ``prepare'' a quantum state, we usually start by isolating and cooling a physical system, which results in an isolated ground state - the $T\to 0$ limit of Eq.(\ref{eq:thermal}). We can then obtain the desired state by applying fields, the effects of which are described by unitary transformations defined by an action $S_n$ as shown in Eq.(\ref{eq:unitary}). A quantum state provides a mathematical summary of these processes, which should allow us to understand the observable effects of our ``preparation'' in interactions with other objects in the laboratory. This is the point where quantum theory causes the most misunderstandings. Firstly, the mathematical description is so abstract that we usually fail to see the relation with the actual physics of quantum state preparation. Secondly, the description of the measurement process is also given in abstract terms, making it impossible to identify the outcomes of measurements with ``elements of reality''. The latter problem is well known and has led to the controversies about different interpretations of quantum mechanics. What we can say for sure is that the quantum state is not a conventional description of physical reality, since it does not describe the system by assigning precise values to the observable properties of the object. Likewise, quantum measurement theory does not provide us with a conventional description of causality, where the measurement outcome is simply an effect caused by a well-defined property of the object. 

The standard textbook solution to the measurement problem is to assume that a precise measurement of a physical property $\hat{A}$ will result in an outcome given by an eigenvalue $A_a$ of the self-adjoint operator $\hat{A}$, where the probability of the outcome is given by the projection on the eigenstate $\mid a \rangle$ of the operator. The problem with this approach is that it only applies to a very narrow range of measurements, and these kinds of measurements are not really representative of physics in general. Thus, the measurement postulate fails to connect the description of physical properties by self-adjoint operators with the experimental reality of physics in the laboratory. A proper understanding of both state preparation and measurement requires a closer look at the physics that is being summarized by the mathematical expressions. The question is whether the Hilbert space formalism itself already gives us some clues about the relations between the physics of state preparation and measurement on the one side, and the mathematics of state vectors and projectors on the other. Based on recent research, I would say that the essential insight is contained in the representation of dynamics by unitary transformations, as represented by the relation between Eqs.(\ref{eq:thermal}) and (\ref{eq:unitary}).

\section{Quantum ergodicity}
\label{sec:ergodic}

Let us start with the problem of state preparation. The starting point is a cooling process which involves random interactions that have no specific time dependence. As a result, the system is left in a completely random phase of its motion, which is why thermal statistics can be derived using the ergodic hypothesis that identifies ensemble averages with time averages. In quantum mechanics, state preparation most often starts from an energetic ground state. However, the Hilbert space formalism makes no fundamental distinction between ground states and excited states. Motion is described by Eq.(\ref{eq:unitary}), and in that equation, energy eigenstates are stationary because they represent ergodic averages over the motion described by $\hat{U}(t)$. This fact can be confirmed by considering an alternative method of state preparation, where an arbitrary physical property $\hat{A}$ is determined by a precise measurement. This requires an interaction that conserves $\hat{A}$ while changing all other properties according to a random force $\phi$ that represents the back-action of the meter on the system. The effect on an arbitrary initial state $\rho(\mbox{in})$ is given by
\begin{eqnarray}
\label{eq:dephase}
\rho(\mbox{out}) = \lim_{L \to \infty} \frac{1}{L} \int_0^L \exp(-i \frac{\phi}{\hbar} \hat{A})  & \;\; \hat{\rho}(\mbox{in}) \;\; & \exp(i \frac{\phi}{\hbar} \hat{A}) 
\nonumber \\
= \hspace{0.5cm} \sum_a \mid a \rangle \langle a \mid  & \;\; \hat{\rho}(\mbox{in}) \;\; & \mid a \rangle \langle a \mid.
\end{eqnarray}
Thus the pre-condition of a preparation by measurement is a randomization of the dynamics along the trajectory represented by $\mid a \rangle\langle a \mid$. The loss of coherence between different eigenstates finds its physical meaning in the randomization of the dynamics along $a$. We should therefore not think of quantum states as representations of the physical quantity $A_a$ given by the eigenvalue of $\hat{A}$, but as complete orbits of the dynamics generated by the physical property $\hat{A}$. This is precisely why the action plays such a fundamental role in quantum physics. 

The important message here is that state preparation is not just ``knowledge of the property $A_a$.'' The quantum state also contains a memory of the dynamics by which $A_a$ was determined. That is the fundamental reason why we cannot just add information about a different physical property $\hat{B}$ to an initial state $\mid a \rangle \langle a \mid$. The orbit $\mid b \rangle \langle b \mid$ is fundamentally different from the orbit $\mid a \rangle \langle a \mid$, and there is no ``joint orbit'' of $a$ and $b$. Nevertheless, there is a kind of intersection between the two orbits, and this intersection obtains its physical meaning when a precise measurement of $\hat{B}$ is performed after the preparation of $\mid a \rangle \langle a \mid$. Specifically, the measurement is just the time reverse of a quantum state preparation, and the reason why the outcome $B_b$ should not be mistaken for a measurement independent ``element of reality'' is that it can only be obtained after the system was driven through the complete orbit described by $\mid b \rangle\langle b \mid$. Note that this observation is closely related to the role of the eigenvalues in the dynamics generated by an operator. The original motivation for the formulation of Eq.(\ref{eq:unitary}) was that the frequencies of dipole oscillations in atomic transitions correspond to differences between the energy levels. In other words, the differences between energy eigenvalues $E_n-E_m$ correspond to periodicities $T$ in the dynamics of the system,
\begin{equation}
E_n-E_m = \frac{2 \pi \hbar}{T_{nm}}.
\end{equation}
Importantly, $T_{nm}$ is a property of the complete orbit generated by the operator of energy $\hat{H}$. Therefore, the energy eigenvalues $E_n$ cannot just represent the energy of a single point along the orbit, but need to be associated with the dynamics of the entire orbit. Experimental observation of quantized values necessarily require interactions that sample the complete orbit generated by the observable. Physical effects that do not involve a complete orbit cannot resolve a specific eigenvalue. The emergence of quantized eigenvalues is therefore a feature of the dynamics, and not a static reality of the non-interacting system. 

We can now get a better physical understanding of the textbook version of a quantum measurement by considering the relation between the initial randomization of the dynamics represented by $\mid a \rangle \langle a \mid$ and the final randomization represented by $\mid b \rangle \langle b \mid$. The measurement outcome $b$ is obtained from $a$ because the two orbits intersect, and the statistical weight of the intersection between the orbits, which corresponds to the dwell time of $a$ in $b$ (or equivalently, of $b$ in $a$), is given by the well known formula
\begin{equation}
P(b|a) = \mbox{Tr}\left(\mid b \rangle \langle b \mid a \rangle \langle a \mid\right).
\end{equation}
This standard rule of quantum statistics therefore represents a relation between the dynamics along $a$ and the dynamics along $b$, which corresponds to the classical phase space geometry of ergodic orbits.

In general, a quantum system will also evolve in time between the initial preparation and the final measurement, so it may be useful to take a closer look at the way that the unitary transformation in Eq.(\ref{eq:unitary}) connects state preparation and measurement when the operators $\hat{A}$, $\hat{B}$ and $\hat{H}$ do not commute. In that case, the eigenstates $\mid a \rangle$ and $\mid b \rangle$ can be represented by superpositions of the eigenstates of energy, and the time dependent probability of finding $a$ after a time $t$ is 
\begin{eqnarray}
\label{eq:time}
P(b|a;t) &=&  \langle b \mid \hat{U}(t) \mid a \rangle \langle a \mid \hat{U}^\dagger(t) \mid b \rangle
\nonumber \\ &=&
\left|\sum_n \exp(-i\frac{S_n}{\hbar}) \langle b \mid \psi_n \rangle \langle \psi_n \mid a \rangle \right|^2.
\end{eqnarray}
In this context, it is interesting to consider how much time it will take to get from $a$ to $b$. In quantum mechanics, this is a somewhat ambiguous question, since we can only determine the probability of $b$ at a time $t$. A meaningful answer is only obtained if the superposition of energy eigenstates in $a$ results in a highly localized peak in the time dependence of this probability. It is therefore more useful to ask at what time the probability of $b$ is maximal for an initial state $\mid \phi(a) \rangle$ centered around $a$ and an average energy of $E$. For such a localized state, the maximal probability of $b$ is reached when the time evolution of the phases in Eq.(\ref{eq:time}) cancels out the phase differences that exist at $t=0$. We can therefore conclude that the transformation distance between $a$ and $b$ is given by the energy dependent quantum phases, which can be evaluated in terms of the energy dependent action
\begin{equation}
\label{eq:tdist}
S_n (\mbox{max.}) = \hbar \; \mbox{Arg}(\langle b \mid \psi_n \rangle \langle \psi_n \mid a \rangle).
\end{equation}
This relation shows that the phases in the eigenstate decompositions of $\mid a \rangle$ and $\mid b \rangle$ have a clear physical meaning: they describe the transformation distance between $a$ and $b$ along the orbits $\psi_n$. It is possible to connect this to the classical notion of a transformation distance as the time $t$ needed to get from $a$ to $b$ along an orbit of specific energy $E$. In Eq.(\ref{eq:time}), the action of the time evolution is given by $S_n = E_n t$. Phases in the vicinity of $E_n$ are equal when the energy gradient of $S_n (\mbox{max.})$ is compensated by the gradient of $S_n = E_n t$, which is given by the time $t$. For that purpose, we can approximate the action $S_n (\mbox{max.})$ by a continuous function of energy $S(E)$, where the continuous energy $E$ represents the expectation value of a minimum uncertainty state centered around $a$ and $E$. We can then use the energy dependence of the quantum mechanical action phase $S_n(\mbox{max.})$ to determine the classical limit of the time of propagation between $a$ and $b$ at energy $E$,
\begin{equation}
t(b,a,E) = \frac{\partial}{\partial E} S(b,a,E).
\end{equation}
As discussed in more detail in \cite{Hof16}, we can use this relation to derive quantum mechanical phases directly from the classical description of the dynamics. In fact, the notion of transformation distance allows us to derive quantum interference effects from the classical laws of dynamics, which provides a physical explanation of the main differences between quantum statistics and classical phase space statistics. Specifically, it is possible to derive a phase space analog of quantum statistics that incorporates the transformation distance in the form of complex phases for the joint and conditional probabilities that relate non-commuting physical properties to each other.

\section{Phase space analogs and their limitations}
\label{sec:phasespace}

The identification of projection operators with orbits raises an important question about the physics of Hilbert space. Why is it that the orbits cannot be expressed as a sequence of points that correspond to the changing values of physical properties along the orbit? Why is it that the intersection between two orbits does not identify a phase space point defined by the pair of eigenvalues that characterize the two orbits? We can actually use the concept of transformation distance to address this question. 

Classical phase space points provide a compact description of all physical properties. For example, the intersection of the orbits $a$ and $b$ would provide a well defined value for the energy $E$, and this value would be found where the transformation distance between $a$ and $b$ along $E$ was zero,
\begin{equation}
\frac{\partial}{\partial E} S(b,a,E) = 0.
\end{equation}
In quantum mechanics, this relation can be no more than an approximation. If we look at the definition of transformation distance in Eq.(\ref{eq:tdist}), we can see that this approximation relates to a stationary phase in Hilbert space. As shown in \cite{Hof11}, this phase also appears in weak measurements of the probability of finding $E_n$ conditioned by an initial state $a$ and a final state $b$,
\begin{equation}
\label{eq:S}
S(b,a,E_n) = \hbar \mbox{Arg}\left(\frac{\langle b \mid \psi_n \rangle \langle \psi_n \mid a \rangle}{\langle b \mid a \rangle}\right).
\end{equation}
It is interesting to note that coarse graining this complex weak value over an energy interval $\Delta E$ will eliminate contributions with action derivatives much greater than $\hbar/\Delta E$, leaving only results in the vicinity of the classical solution $E(b,a)$ \cite{Hof11,Hof12,Hof14a}. This means that weak values establish a physically meaningful link between Hilbert space and classical phase space. What is even more astonishing is that the mathematics of this phase space analogy was already discovered in the early days of quantum mechanics, when it was constructed from the operator algebra as an alternative to the Wigner distribution \cite{McCoy32,Kir33,Dir45}. Unfortunately, these mathematical insights were mostly forgotten by the time that Aharonov, Albert and Vaidman introduced weak measurements and their result, the weak values \cite{Aha}. It was therefore not immediately recognized that the oddities of weak values merely describe the differences between classical phase space concepts and their more accurate description in the Hilbert space formalism. However, recent experimental demonstrations have shown that weak measurements can be used to directly measure quantum states as weak joint probabilities of two non-commuting observables \cite{Lun11,Lun12,Wu13,Sal13,Bam14}. 
These results show that weak values represent the quantum mechanical analog of classical phase space statistics, including the non-classical correlations between physical properties that cannot be measured jointly. It is also worth noting that weak values can also be observed at finite measurement strengths, indicating that the algebra of weak values provides an experimentally relevant description of non-classical correlations \cite{Suz12,Hof12b,Hir13,Hof14b,Zou15,Kin15,Val16,Iin16}. In fact, many of the recent experimental investigations of quantum paradoxes have used weak measurements to show that the paradoxical features can be understood as a direct consequence of the negative weak conditional probabilities associated with action phases of $\hbar \pi$ in Eq.(\ref{eq:S}) \cite{Res04,Jor06,Tol07,Wil08,Lun09,Yok09,Gog11,Suz12,Den14,Hof15}. 

With this large number of results from both experiment and theory, it is rather surprising that so little attention has been paid to the role that the operator algebra plays in determining the non-classical statistics that are observed by weak measurements and related methods. As shown in \cite{Hof14c}, it is actually possible to argue that the Hilbert space algebra itself defines the ordered product of the projection operators as the only reasonable representation of joint probabilities for the possible measurement outcomes of two non-commuting observables. We can now understand this result in terms of the identification between projectors and orbits discussed above. The joint statistical weights of two orbits in a quantum state $\hat{\rho}$ are then given by
\begin{eqnarray}
\rho(a,b) &=& \langle \mid b \rangle\langle b \mid a \rangle\langle a \mid \rangle
\nonumber \\
&=& \langle b \mid a \rangle \langle a \mid \hat{\rho} \mid b \rangle.
\end{eqnarray}
It is fairly easy to see that this is a complete description of the state $\hat{\rho}$ for any two basis systems with non-zero mutual overlaps $\langle b \mid a \rangle$. In fact, this expression was already introduced as a phase space analog by Dirac in 1945, and is therefore often referred to as the Dirac distribution \cite{Dir45}. In the same work, Dirac also introduced weak values as a mathematical description of operators. In terms of the operator algebra, we can see that the weak values for all combinations of $a$ and $b$ give a complete description of the operator $\hat{M}$ as
\begin{equation}
\label{eq:wv}
\hat{M} = \sum_{a,b} \frac{\langle b \mid \hat{M} \mid a \rangle}{\langle b \mid a \rangle}
\; \mid b \rangle \langle b \mid a \rangle \langle a \mid.
\end{equation}
Thus weak values are closely associated with the idea that the product of projection operators represents the intersection of two orbits and therefore corresponds to the closest analogy to a phase space point that can be defined in quantum physics. 

A complete description of the operator algebra associated with complex joint probabilities has been given in \cite{Hof12}. For the present purpose, it is sufficient to note that the strangeness of the statistics associated with weak values and complex probabilities arises from the dynamical relations between the physical properties. It is therefore not possible to assign an eigenstate $\mid m \rangle$ of the operator $\hat{M}$ to the combination of orbits $(a,b)$. Instead, the contribution of the orbit $m$ to the intersection of the orbits $a$ and $b$ is given by a complex conditional probability,
\begin{equation}
P(m|a,b) = \frac{\langle b \mid m \rangle \langle m \mid a \rangle}{\langle b \mid a \rangle},
\end{equation}
where the probability $P(m)$ of finding $m$ for a specific Dirac distribution $\rho(a,b)$ is given by the standard form for conditional probabilities,
\begin{equation}
P(m)= \sum_{a,b} P(m|a,b) \rho(a,b).
\end{equation}
As shown in Eqs.(\ref{eq:tdist}) and (\ref{eq:S}), the complex conditional probability $P(m|a,b)$ describes transformation distances between the different orbits rather than joint realities \cite{Hof11,Hof14a}. It is therefore necessary to distinguish the reality of a precise measurement outcome from the dynamical relations between physical properties. 

\section{The relation between mathematics and physical reality}
\label{sec:measurement}

We now turn to the central question that has caused so much confusion in quantum physics. How do the physical properties of a system appear in the outcomes of an actual experiment? As mentioned at the end of Sec. \ref{sec:physics}, the concept of measurement given by most textbooks of quantum mechanics is actually too narrow to accommodate all of the possible interactions of a physical system. In a more general description of measurements, the outcome $m$ is represented by an operator $\hat{E}_m$, so that the probability of obtaining $m$ for a quantum state $\mid \psi \rangle$ is given by 
\begin{equation}
P(m|\psi) = \langle \psi \mid \hat{E}_m \mid \psi \rangle.
\end{equation}
Effectively, the operator $\hat{E}_m$ describes the conditional probability of obtaining $m$ for arbitrary initial conditions $\mid \psi \rangle$. But what is the relation between the measurement outcome $m$ and the physical properties of the system? The specific realization of the measurement should give a non-trivial answer, and that answer must somehow enter into the Hilbert space description as well. 

As mentioned in the introduction above, an interesting solution to the problem was presented by Ozawa \cite{Oza03} and has recently been investigated in a number of experiments \cite{Has12,Roz12,Bae13,Wes13,Rin14}. In this approach, the operator $\hat{A}$ is used to represent the target observable, and a quantitative estimate $\tilde{A}_m$ is associated with each measurement outcome $m$. The measurement error is then given by the difference between the operator $\hat{A}$ and the value $\tilde{A}_m$. Since this expression is itself an operator, it needs to be evaluated using the operator algebra. The expression for the total measurement error derived by Ozawa is
\begin{equation}
\label{eq:Oza}
\epsilon^2(A) = \sum_m \langle \psi \mid (\tilde{A}_m-\hat{A}) \hat{E}_m (\tilde{A}_m-\hat{A}) \mid \psi \rangle.
\end{equation}
As we discuss in a recent paper \cite{Nii16}, this relation makes a non-trivial statement about the relation between quantitative properties and measurement outcomes. Specifically, we show that the only possible definition of joint statistical weights for the eigenstate outcomes $a$ and the actual outcomes $m$ that is consistent with the quantitative definition of the error in Eq.(\ref{eq:Oza}) is given by 
\begin{equation}
P(m,a|\psi) = \mbox{Re}(\langle \psi \mid \hat{E}_m \mid a \rangle\langle a \mid \psi \rangle).
\end{equation}
Since the algebra of Hilbert space corresponds directly to the algebra of classical probabilities, the optimal estimate is then given by the real part of the weak value of $\hat{A}$, as already pointed out by Hall in \cite{Hal04}, soon after the initial concept had been introduced by Ozawa.  

In the present context, it is important to realize that the weak values are optimal estimates because they accurately summarize the causality relations between non-commuting physical properties in the Hilbert space formalism. To understand the relation between measurement outcomes and causality better, one should keep in mind that the initial state $\mid \psi \rangle$ represents an orbit generated by a specific physical property $\hat{B}$, so that $\mid \psi \rangle$ is an eigenstate of $\hat{B}$ with an eigenvalue of $B_\psi$. We can now add the quantity $\hat{A}$ to the value of $\hat{B}$ to obtain a new quantity $\hat{M}$, and this quantity defines a new set of orbits $\mid m \rangle$. A quantitative measurement of $\hat{M}$ identifies the eigenvalue $M_m$ of the final orbit. Since the quantity $\hat{A}$ is defined as the difference between $\hat{M}$ and $\hat{B}$, it is clear that its value should be
\begin{equation}
\label{eq:determine}
A(\psi,m) = M_m - B_\psi.
\end{equation}
Here, the quantity $A(\psi,m)$ does not refer to an orbit generated by $\hat{A}$. Instead, it related the orbits expressed by $\mid \psi \rangle$ and $\mid m \rangle$ to each other. Specifically, $M_m$ and $B_\psi$ are eigenstates of $\mid m \rangle$ and $\mid \psi \rangle$ for operators $\hat{M}$ and $\hat{B}$, defined in such a way that $\hat{A}$ can be expressed as the operator sum $\hat{M}+\hat{B}$ as shown in \cite{Nii16},
\begin{eqnarray}
A(\psi,m) &=& \frac{\langle m \mid (\hat{M} + \hat{B}) \mid \psi \rangle}{\langle m \mid \psi \rangle}
\\ \nonumber
&=& \frac{\langle m \mid \hat{A} \mid \psi \rangle}{\langle m \mid \psi \rangle}.
\end{eqnarray}
Note that in the general case the values of $A(\psi,m)$ are complex, requiring a non-hermitian operator $\hat{M}$ for the assignment of complex values $m_m$ to the measurement outcomes $m$. However, such an assignment is not necessarily meaningless since the purpose of the present analysis is to identify a precise relation between the value $A(\psi,m)$ of $\hat{A}$ and the eigenvalues $M_m$ and $B_\psi$, where the statistical errors in the quantitative relation are zero. Ozawa's error relation confirms this expectation by defining the contribution of $m$ to the error $\epsilon^2(A)$ as
\begin{eqnarray}
\epsilon(A,m) &=& \langle m \mid(\hat{A} - \tilde{A}_m) \mid \psi \rangle 
\nonumber \\
&=& \left( A(\psi,m) - \tilde{A}_m \right) \langle m \mid \psi \rangle. 
\end{eqnarray}
This contibution is zero whenever the estimate $\tilde{A}_m$ is equal to the complex weak value conditioned by $\psi$ and $m$. In the example above, $\tilde{A}_m = A(\psi,m)$ is only possible when the weak value $A(\psi,m)$ is real, so that a measurement error of $\epsilon^2(A)=0$ is only possible when all of the weak values associated with different measurement outcomes $m$ are real. 
However, there is no logically binding reason to maintain the restriction to real values when the untimate goal is the identification of deterministic relations between physical properties. As discussed above, the complex weak value is a valid quantification of the intersection of the orbits $\mid \psi \rangle$ and $\mid m \rangle$ in terms of the quantity defined by the operator $\hat{A}$. By extending the estimate to complex values, it is always possible to obtain the error free value $\tilde{A}_m = A(\psi,m)$ from a maximally precise measurement. Since the value $A(\psi,m)$ is error free, it can serve as a deterministic expression of the relation between the value of $A$ and the precisely defined conditions $\psi$ and $m$ which holds for all quantum states $\hat{\rho}$. We can verify that this is indeed correct by using the joint statistics of $\psi$ and $m$ defined by the Dirac distribution of $\hat{\rho}$,
\begin{equation}
\rho(m,\psi) = \langle \psi \mid \hat{\rho} \mid m \rangle \langle m \mid \psi \rangle.
\end{equation}
Note that here, the quantum state is given by $\hat{\rho}$, whereas $\psi$ is merely a basis state used to characterize the statistics of $\hat{\rho}$. The expectation value of $\hat{A}$ in $\hat{\rho}$ can now be explained as an average of the deterministic values $A(\psi,m)$ of $\hat{A}$ determined by the combinations of $\psi$ and $m$,
\begin{equation}
\langle \hat{A} \rangle = \sum_{m,\psi} A(\psi,m) \rho(m,\psi). 
\end{equation}
We can therefore conclude that error free relations between $\hat{A}$, $m$ and $\psi$ provide a state independent description of deterministic relations between physical properties \cite{Hof12,Hof14a}.

\section{Empirical objectivity and non-classical correlations}
\label{sec:objective}

The central merit of the Hilbert space formalism is that it provides an objective description of the quantum system. In quantum mechanics, this presents a problem because we cannot simply neglect the role of the environment in the physical processes used to prepare and measure the system. In popular discussions of quantum physics, it has often been suggested that quantum physics involves some mysterious influence of the observer on the result, implying the complete absence of objective laws of causality. It is therefore important to stress that the Hilbert space formalism does not allow any such ``external'' effects. Even the description of preparation and measurement is entirely objective. The problem arises only from the possible choice between different state preparations or measurement procedures. However, these procedures are all defined by physical interactions with the object, and the effects of these physical interactions can then be described objectively by using the Hilbert space algebra. 

We need to understand the algebra of Hilbert space as a description of causality that relates a physical object to the evidence of its existence found outside of the system. Objectivity is only possible if we can apply rules of causality to eliminate the unavoidable contextuality introduced by external devices. Quantum physics shows that the most fundamental elements of reality are processes, not properties. Processes can be objectified as orbits described by Hilbert space projectors. The result is a proper causal description of the system, where the self-adjoint operators describing physical properties can be used to evaluate the quantitative effects observed at finite sensitivities. 

It is somewhat unfortunate that quantum physics is rarely applied properly to systems that behave in a nearly classical fashion. It is important to remember that the classical description of such systems is merely approximate, no matter how large they are. In most cases, the observation of objects involves fluctuations that are much larger than the quantum limit. Just as an extreme example, we can consider the motion of the moon around the earth. At a distance of about 400 000 km from the center of the earth, a single photon of visible light scattered by the surface of the moon will change the angular momentum by about $5 \times 10^{15} \hbar$. Since no classical description of the orbit of the moon can take into account every single photon scattered by the moon, it is obvious that classical physics is no more than a very crude approximation - except by relative standards, of course, where we should consider that the total angular momentum of the moon going around the earth is about $3 \times 10^{68} \hbar$. The motion of the moon is therefore quite robust against the disturbances caused by the light we need to see it by. We should just avoid the misconception that the moon has a reality independent of its interaction with light and matter. The fact that the moon is continuously immersed in interactions with its environment makes the moon real, just as all other objects are only real as a source of their physical interactions. 

The detailed investigations of non-classical correlations we have recently performed indicate that we should take imaginary correlations seriously \cite{Kin15,Iin16}. This is a direct consequence of the relation between unitary transformations and statistics in the operator algebra. Specifically, the imaginary correlations of two non-commuting operators $\hat{A}$ and $\hat{B}$ is given by the expectation value of the commutation relation,
\begin{equation}
\mbox{Im}\left(\langle \hat{A}\hat{B} \rangle\right) = - \frac{i}{2} \langle [\hat{A},\hat{B}] \rangle.
\end{equation}
Therefore, the time evolution of any physical property $\hat{A}$ is evidence of an imaginary correlation between $\hat{A}$ and the energy $\hat{H}$,
\begin{equation}
\label{eq:ncorr}
\mbox{Im}\left(\langle \hat{A}\hat{H} \rangle\right) = \frac{\hbar}{2} \frac{d}{dt} \langle \hat{A} \rangle.
\end{equation}
Importantly, it is possible to experimentally observe the imaginary correlation between $\hat{A}$ and $\hat{H}$ in weak measurements or in any other experimental reconstruction of the Dirac distribution $\rho(A,H)$. Oppositely, it is not possible to observe the change in $\hat{A}$ without changing the energy $\hat{H}$ as a result of the necessary interactions. Therefore, the identification of the rate of change with an imaginary correlation does not contradict our experience. Rather, the assumption that we observe physical properties as exact real numbers is at odds with the empirical evidence. We can quickly confirm that the limit placed by Eq.(\ref{eq:ncorr}) on our ability to estimate both the energy and the value of $\hat{A}$ from the evidence is not unrealistically high. After all, $\hbar$ is a very small action. For example, the imaginary correlation between position and energy achieves its maximal possible value at the speed of light, where it is a mere $1.58 \times 10^{-26}$ Jm. The lesson we should learn from such considerations is that the assumption that we could hypothetically control physical systems with absolute precision is mostly a fantasy based on sloppy thinking. Quantum mechanics reveals that we need to make corrections to the artificially precise laws of motion once we approach the limit where small actions do matter. Nevertheless the laws of motion remain objective and consistent. The origin of the randomness observed in quantum experiments is explained by the limitations of control that these deterministic laws of motion impose on the possible interactions with the system. We can understand these limitations once we realize that the mathematical formalism describes dynamics and causality, and not the static realities represented by the classical phase space algebra.

\vspace{-0.5cm}

\section{Conclusions}
\label{sec:conclusions}

In quantum theory, Hilbert space is used to describe the deterministic relations between physical properties that allow us to trace external effects of a system back to causes within the system. Objective reality emerges as a result of the causality relations between observations of the same object made at different times, or, in the spirit of ergodic theory, between observations made on identically prepared objects of the same type. Importantly, the physical properties of an object are known only through the effects of interactions - by ``touch and sight.'' It is a serious mistake to assume that reality is accessible by abstract thought. The elements of the theory do not represent platonic realities. Each one of them needs to be justified by actual effects observed in the laboratory. This demand may seem overly restrictive, and it should not be taken as an attempt to reject speculations about possible observations that have not been realized due to technical limitations - but the present discussion of quantum mechanics suffers from unnecessary confusion because scientists cling to concepts of reality that are clearly at odds with the observed phenomena. A more careful distinction between the observable world and unobservable figments of the imagination may therefore be helpful. In particular, we should be more humble in admitting that our knowledge of reality is limited to our actual experience, and that the extrapolations of our personal experience to possible experiences beyond our technical capabilities may result in delusions about the real world. Science should provide a tool by which we can reach an agreement on questions about the external reality, and this can only be achieved if there is a shared experience of the world that we can all relate to. 

The intention of the analysis of quantum theory developed here and in a number of related works \cite{Hof16,Hof12,Hof14a} is to provide an empirical foundation of quantum physics that explains how the formalism describes the observable laws of physics that shape our experience of the world around us. At the center is the realization that objects obtain their reality by their appearance, and the properties of the object are the quantities that determine the possible effects of the object that determine its appearance, both in the laboratory and in nature. The abstraction of the ``state'' should really be understood in terms of this experience, where the projection on a Hilbert space vector actually represents an interaction that randomizes the dynamics of the system in the course of the interaction by which the object causes an observable effect. The strangeness of quantum statistics originates from the peculiar role played by the laws of motion that determine this dynamical randomization. Specifically, our approximate separation between reality and dynamics - the assumption of a static reality - breaks down when the interaction is sensitive to actions of $\hbar$ or less. This is no different from the breakdown of the independence of time and motion when velocities approach the speed of light. It may therefore be possible to gain a better fundamental understanding of physics by noticing that nothing in our experience indicates that the reality of objects is static and can be frozen in time. Quantum mechanics simply shows that this unnecessary assumption is wrong, and that dynamics forms an essential part of objective reality in the limit of small actions.

\end{document}